\documentclass{article}

\usepackage{PRIMEarxiv}

\usepackage[utf8]{inputenc} %
\usepackage[T1]{fontenc}    %
\usepackage{hyperref}       %
\usepackage{url}            %
\usepackage{booktabs}       %
\usepackage{amsfonts}       %
\usepackage{nicefrac}       %
\usepackage{microtype}      %
\usepackage{lipsum}
\usepackage{caption}
\usepackage{subcaption}
\usepackage[table,xcdraw]{xcolor}

\usepackage{wrapfig}
\usepackage{fancyhdr}       %
\usepackage{graphicx}       %
\usepackage{float}
\usepackage{amsmath}

\graphicspath{{figures/}}     %

\usepackage{placeins}
\usepackage{soul,color}
\usepackage{multirow}
\title{ Probabilistic Generative Transformer Language models for Generative Design of Molecules
\thanks{\textit{\underline{Citation}}: 
\textbf{L.W...J.H. . Probabilistic Generative Transformer Language models for Generative Design of Molecules. DOI:000000/11111.}} 
}

\author{
 Lai Wei\\
 Department of Computer Science and Engineering\\
  University of South Carolina\\
  Columbia, SC 29201 \\
  \And
 Nihang Fu, Yuqi Song \\
 Department of Computer Science and Engineering\\
  University of South Carolina\\
  Columbia, SC 29201 \\  
   \And
 Qian Wang\\
 Department of Chemistry and Biochemistry\\
  University of South Carolina\\
  Columbia, SC 29201 \\    
   \And
 Jianjun Hu *\\
 Department of Computer Science and Engineering\\
  University of South Carolina\\
  Columbia, SC 29201 \\
  \texttt{jianjunh@cse.sc.edu} \\
}

\begin{document}
\maketitle

\begin{abstract}

Self-supervised neural language models have recently found wide applications in generative design of organic molecules and protein sequences as well as representation learning for downstream structure classification and functional prediction. However, most of the existing deep learning models for molecule design usually require a big dataset and have a black-box architecture, which makes it difficult to interpret their design logic. Here we propose Generative Molecular Transformer (GMTransformer), a probabilistic neural network model for generative design of molecules. Our model is built on the blank filling language model originally developed for text processing, which has demonstrated unique advantages in learning the "molecules grammars" with high-quality generation, interpretability, and data efficiency. Benchmarked on the MOSES datasets, our models achieve high novelty and Scaf compared to other baselines. The probabilistic generation steps have the potential in tinkering molecule design due to their capability of recommending how to modify existing molecules with explanation, guided by the learned implicit molecule chemistry. The source code and datasets can be accessed freely at \url{https://github.com/usccolumbia/GMTransformer}

\end{abstract}


\section{Introduction}

The discovery of novel organic molecules has wide applications in many appliations such as drug design and catalysis development\cite{meyers2021novo}. However, due to the sophisticated structure-property relationships, traditional rational design approaches have only covered an extremely limited chemical design space\cite{zunger2021understanding}. 
Recently, a large number of generative machine learning algorithms and models have been proposed for molecule design as systematically reviewed in \cite{meyers2021novo,du2022molgensurvey,imrie2020deep}. The first category of these methods are deep generative models mainly including Variational Autoencoders (VAEs)\cite{dai2018syntax}, Generative Adversarial Networks (GANs)\cite{guimaraes2017objective}, and normalizing flow-based models \cite{zang2020moflow}. Two of the major limitations of these models includes their black-box nature of the models and their challenge to deal with modularity in molecule design. the black-box nature of the deep neural network-based generator makes it difficult to interpret the black-box GAN models in terms of the chemical knowledge they learn and how they exploit the learned implicit knowledge for a generation.
The second category of molecule generative design methods include several key combinatorial optimization algorithms such as genetic algorithms\cite{kwon2021evolutionary}, reinforcement learning\cite{blaschke2020reinvent}, Bayesian optimization\cite{winter2019efficient}, Monte Carlo Tree Search (MCTS)\cite{yang2017chemts}, Markov Chain Monte Carlo (MCMC) \cite{du2022molgensurvey}. While GAs have demonstrated superior performance in several molecule design benchmark studies \cite{huang2021therapeutics,brown2019guacamol}, the genetic operators of mutation and cross-over lack the learning capability to achieve intelligent and efficient chemical space exploration. This also applies to MCTS, which locally and randomly searches each branch of intermediates and selects the most promising ones during each generation iteration \cite{yang2020practical}. Bayesian optimization is usually applied together with VAEs and search the chemical space in the latent space, which makes it difficult to handle the chemical constraints explicitly \cite{winter2019efficient} and also cannot handle modularity in molecule design. Reinforcement learning has been applied to generative models with both SMILES and 2D graph representations, which learns a policy network to determine the optimal actions that maximize a global reward such as a given property\cite{zhou2019optimization, blaschke2020reinvent}. However, RL is rarely used in de novo molecule generation partially due to the difficulty to achieve long-range credit assignment and to obtain differentiable validity check as the reward signal.

Another important consideration in the design of generative models for molecules is the representation level of the molecules, which includes atom based, fragment based and reaction based approaches. Most of existing models have used a single the atom based representations such as SMILES \cite{weininger1988smiles} while other more advanced representations such as SELFIES \cite{krenn2019selfies} and DeepSMILES \cite{o2018deepsmiles} have been proposed for molecule property prediction. How the choice of the molecule representation affects the generative design performance remains unsettled. It is also found that the basic atom representations such as SMILES are not easy to be used to exploit the modules, motifs, or skeletons of known molecules. On the other hand, while fragment and reaction based generative models can exploit such larger building blocks, they also have an issue in their expression power.

Another major limitation of existing deep generative models for molecule design is that most of them cannot be used for tinkering design: a specified part of an existing molecule is masked for replacement of other modules to gain specific function property, despite that this is one of the most widely used approaches to explore new molecules \cite{zunger2021understanding} due to many constraints imposed on the possible options. During these processes, chemists or molecules scientists usually resort to their intuition, chemical knowledge, and expertise to select substitution or doping elements and proportions to tune the properties of the molecule by considering a variety of factors such as chemical compatibility, poison level, geometric compatibility, synthesizabilty, and other heuristic knowledge.

Here we propose a self-supervised probabilistic language model Generative Molecular Transformer (GMTransformer) for molecular design and generation. The model is based on transformers and self-supervised blank-filling language model BLM \cite{shen2020blank}. The model interpretably calculates its probabilities and derives different actions depending on the token frequency shown by its vocabulary. We use SMILES, SELFIES and DeepSMILES representations to train different models, and found that each of them has its own advantage. The easy interpretation, data efficiency, and tinkering design potentials have been demonstrated in our recent work on inorganic materials composition design \cite{wei2022crystal}, which inspires us to explore its potential in molecule design in this work.
We use MOSES benchmarking metrics to evaluate the performance of our GMTransformer models. The results of our extensive experiments show strong performance compared to the state-of-the-art baselines. Our GMTransformer model with SMILES representation achieves 96.83\% novelty and 87.01\% of IntDiv, which demonstrates that our model is capable of generating a wide variety of novel molecules. We also train generative models for maximizing different properties: logP, tPSA, and QED, and find that our models can learn to generate molecules with specific property as demonstrated by the distribution of generated molecular properties.

\section{Results}
\label{sec:headings}

\subsection{Generative and tinkering molecular design as a blank-filling process}

SMILES (Simplified molecular input line entry system) uses a string of characters to describe a three-dimensional chemical structure. Atom, bond and branche make up the strings of SMILES. The atoms are represented by their element symbols, e.g. C, N, O, S, F. The atoms in aromatic rings are represented by lowercase letters, such as the lowercase c for aromatic carbon. There are three types of bonds in SMILES: single bonds, double bonds and triple bonds and they are denoted by -, =, \# respectively. Branches are specified by enclosures in parentheses.

\begin{table}[th]
\centering
\caption{Strings of SMILES generated as a canvas rewriting process}
\label{tab:canvas}
\begin{tabular}{lll}
\hline
\multicolumn{3}{c}{ Canvas rewriting with 4 actions: (E, \_E, E\_, \_E\_)}                                            \\ \hline
\multicolumn{1}{l|}{Step t}           & Action  & operation  \\ \hline
\multicolumn{1}{l|}{0. \underline{\$1}}          & \_E\_ &Replace \$1 blank with \_C\_ \\ \hline
\multicolumn{1}{l|}{1. \underline{\$1} C \underline{\$2}}    & E & Replace \$1 blank with C     \\ \hline
\multicolumn{1}{l|}{2. C C \underline{\$1}}     & E\_ & Replace \$1 blank with (\_    \\ \hline
\multicolumn{1}{l|}{3. C C ( \underline{\$1}}   & \_E\_ & Replace \$1 blank with \_O\_    \\ \hline
\multicolumn{1}{l|}{4. C C ( \underline{\$1} O \underline{\$2}}    & \_E & Replace \$1 blank with =     \\ \hline
\multicolumn{1}{l|}{5. C C ( = O \underline{\$1}} & E\_ &Replace \$1 blank with )\_      \\ \hline
\multicolumn{1}{l|}{6. C C ( = O ) \underline{\$1}} & E &Replace \$1 blank with C      \\ \hline
\multicolumn{1}{l|}{7. C C ( = O ) C}   &               &                \\ \hline
\end{tabular}
\end{table}

As shown in Table \ref{tab:canvas}, the following canvas rewriting process shows how the GMTransformer generates the $CC(=O)C$ sequence of the SMILES strings step by step. At the beginning, there is only an initial blank token \underline{\$1} on the canvas, then different candidate tokens and rewriting actions (E, \_E, E\_, \_E\_) are selected by GMTransformer. (1) action E: replace a blank with the element E; (2) action \_E: replace a blank with element E and insert a new blank on its left side, allowing further element insertion; (3) action E\_: replace a blank with element E and insert a new blank on its right side, allowing further element insertion; (4) action \_E\_: replace blank with element E and insert new blanks on both sides \cite{wei2022crystal}. Finally, a string without any blank symbol is generated on the canvas. In Table \ref{tab:canvas}, There is only one initial blank on the canvas in step 0, and it selects action \_E\_ with the element C to get \underline{\$1} C \underline{\$2}. Then it replaces the blank of \underline{\$1} with the element C by action E in the first step. In the second step, the operation is replacing \underline{\$1} blank with branch (\_. Then it chooses action \_E\_ and replaces the blank with element O. In the next two steps, it replaces the blank with bond = and branch )\_ respectively to get canvas C C ( = O ) \underline{\$1}. Finally, it replaces \underline{\$1} with element C.

GMTransformer is different from BERT\cite{devlin2018bert} and XL-Net\cite{yang2019xlnet} as it relies on pre-existing content to learn and generate sequences.  Instead of using the context of a pre-masked word to predict the probability of the masked word, GMTransformer directly chooses the action and then inserts the word that best matches the content it learns at the appropriate position based on the probabilistic dependencies in the generated vocabulary.

\subsection{Generative Molecular Transformer: Blank filling language model for molecule generation }

GMTransformer is not like the black-box models such as Variational Autoencoders (VAEs) \cite{dai2018syntax}, Generative Adversarial Networks (GANs) \cite{guimaraes2017objective}, and normalizing flow-based models \cite{zang2020moflow}, it is a process interpretable model designed based on a blank language model (BLM)  \cite{shen2020blank}. GMTransformer directly models the probability of the tokens in the vocabulary. It relies on the content of the existing canvas to calculate the probability distribution to select actions and tokens to generate a new canvas. It can intelligently control the intermediate process of generating the string, and each step can give a explanation of why it is doing it. 

GMTransformer uses SMILES, SELFIES and DeepSMILES representation of atom-level tokenization. The SMILES representation of atom-level tokenization has 21 tokens in SMILES strings and 7 special tokens as the vocabulary during training process. The vocabulary contains 13 atom tokens $<C>, <c>, <O>, <o>, <N>, <n>, <F>, <S>, <s>, <Cl>, <Br>, <[nH]>$, and $<[H] >$, 3 bond tokens $<->, <=>, <\#>$, 6 ring tokens $<1>, <2>, <3>, <4>, <5>, <6>$ and 7 special tokens $<PAD>, <UNK>, <FIRST>, <LAST>, <EOS>, <BLANK>, <BLANK\_0>$. SELFIES and DeepSMILES also contain the same 7 special tokens as SMILES.

The SPE tokenization has a mean length of approximately 6 tokens, while the atom-level tokenization has a mean length of approximate 40. SMILES Pair Encoding contains the special tokens and unique tokens from the frequent SMILES substrings. e.g, $<CCC(C)(C)>$, $<CCCC(C)>$, $<NC(=O)C>$. Both SMILES and DEEP SMILES use the SPE tokenization, which does not apply to SELFIES. More details can be found in \cite{li2021smiles}.

\begin{figure}[ht]
  \centering
  \includegraphics[width=0.8\linewidth]{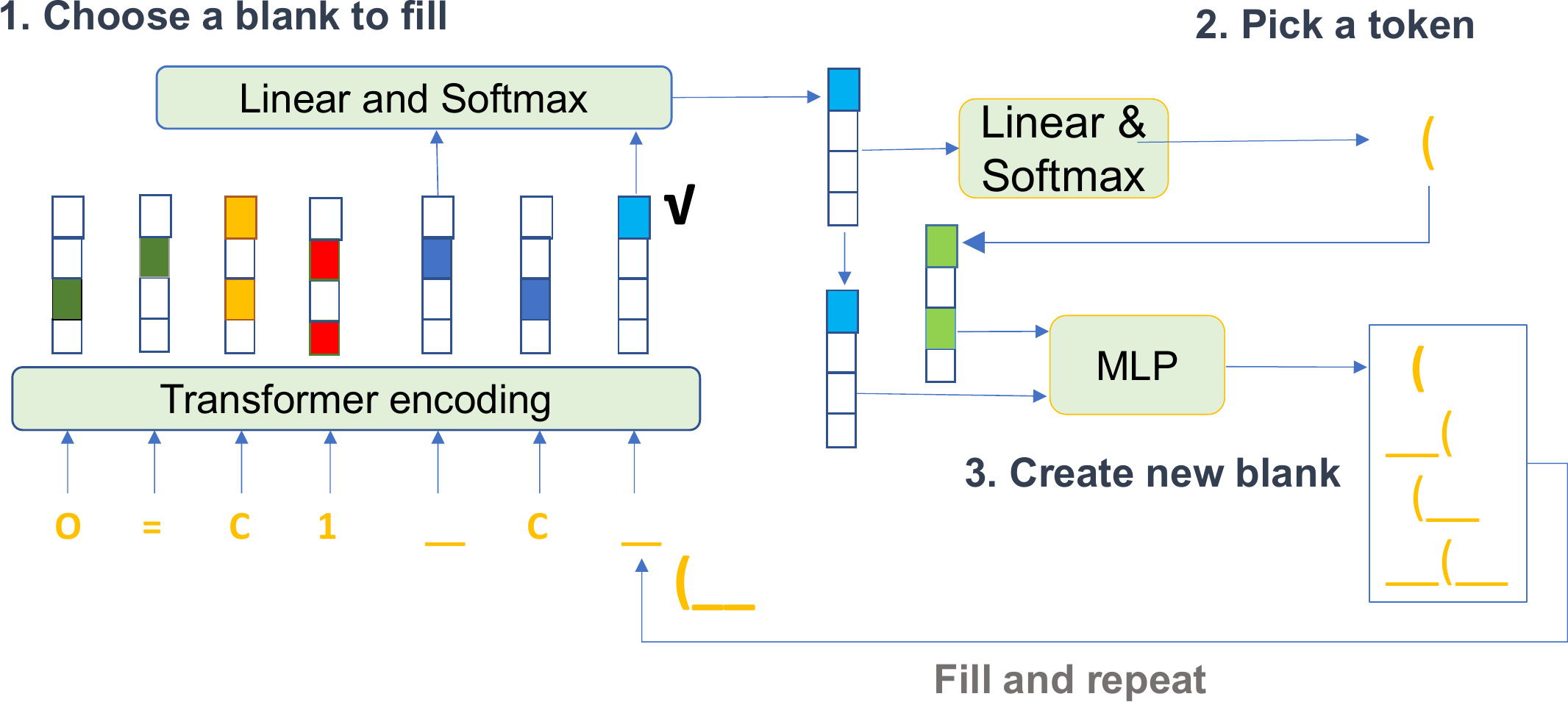}
  \caption{Neural network architecture of the blank filling language model for molecules tinkering using SMILES string $O=C1CC(c2ccccc2)Oc2cc(O)cc(O)c21$ as an example.  }
  \label{fig:architecture}
\end{figure}

Figure \ref{fig:architecture} shows the architecture of our Generative Molecular Transformer (GMTransformer). The model utilizes four networks in three iterable stages. The first stage includes the transformer network and linear and softmax layers. The second and third stages include linear and softmax layers, multi-layer perceptron network, respectively. In the first stage, the transformer network encodes the canvas into a sequence of representations. Then which location of the blank should be filled is selected by computing probabilities from linear and softmax layers. In the second stage, it picks an appropriate token into the blank with linear and softmax layers. In the final stage, the action of whether or not to create blanks to the left and right is determined by feeding the concatenation of the representations of the selected blank and the token into the multi-layer perceptron network. The model updates the canvas and repeats the process until there are no blank positions on the canvas. 

During the training process, first of all, it initializes the model parameter $\theta$ and then randomly samples a training example $x=\left(x_{1}, \cdots, x_{n}\right)$ with the length of $n$. Next, it samples the step length $t$ from $0$ to $n - 1$ and the generation order with n-permutation $\sigma$ of the given example. It constructs a canvas $c$ that remains the first $t$ tokens $x_{\sigma_j}$ ($j=1,...,t)$ and collapse the remaining $n-t$ tokens as blanks. Then it takes $n-t$ target actions $a_{j-t}$ for filling $x_{\sigma_j}$ ($j=t+1,...,n)$ into canvas and calculates loss as Eq. \ref{eq:loss}. Finally it updates parameter $\theta$ by gradient descent and repeat the whole process until convergence. More details can be found in \cite{shen2020blank}.

\begin{equation}
-\log (n !)-\frac{n}{n-t} \sum_{\sigma_{t+1}} \log p\left(a_{t}^{x, \sigma} \mid c_{t}^{x, \sigma} ; \theta\right)
\label{eq:loss}
\end{equation}
where the $\theta$ is the the model parameter; $c_{t}^{x, \sigma}$ is the $t$th canvas with the given training example $x$ and the determined generation order (permutation $\sigma$); $a_{t}^{x, \sigma}$ represents the action whether or not to create blanks to the left and right of the predicted token at step $t$ with the order permutation $\sigma$ and the selected blank.

\subsection{De novo generative design of molecules composition }

\paragraph{Training of GMTransformer for hypothetical molecule generation}
We use the MOSES dataset as our benchmark dataset, which is widely used in generative molecular design community. The performance evaluation criteria is derived from the MOSES package, which is also a standard in generator performance evaluation. 

The GMTransformer model was trained and evaluated using the database of MOSES benchmarking platform.
MOSES is a benchmarking platform to standardize the training results of molecule generation models.
Its initial dataset, ZINK Clean Leads, contains about 4.6 million molecules. The final dataset was obtained by filtering molecules containing charged atoms (except C, N, S, O, F, Cl, Br, H); macrocyclic molecules with more than 8 molecules in the ring; medical chemistry filters (MCFs) and PAINS filters. MOSES provides both training and test sets and a set of metrics for assessing the quality and diversity of the generated molecules. We also evaluate the generated samples of three additional properties: the octanol-water partition coefficient (logP), the topological Polar Surface Area (tPSA), and the Quantitative Estimate of Drug-likeness (QED)\cite{bickerton2012quantifying} computed from RDKit \cite{landrum2019rdkit} are used for training the conditional GMTransformer generator.

\FloatBarrier

\subsection*{Evaluation of GMT's molecular generation performance}

We evaluate the performance of our GMTransformer generators and compare with that of the benchmark molecular models using ten evaluation criteria with MOSES metrics including validity, uniqueness (unique@$1k$ and unique@$10k$), internal diversity (IntDiv), filters, novelty, the similarity to a nearest neighbor (SNN), Frechet ChemNet distance (FCD), fragment similarity (Frag), and scaffold similarity (Scaf). 
As shown in \ref{table:performance}, GMT-SMILES, GMT-PE-SMILES and GMT-SELFIES generate 85.87\%, 82.88\% and 100\% valid samples, respectively. The uniqueness of all models is almost 100\%. Especially, the novelty of GMT-SMILES, GMT-PE-SMILES and GMT-SELFIES is as high as 95.31\%, 88.29\% and 96.83\% respectively. At the same time, GMT-SMILES, GMT-PE-SMILES and GMT-SELFIES have the highest values 85.69\%, 85.58\% and 87.01\% of IntDive respectively among all benchmark models. These high values mean that they can generate samples with higher diversity, which may accelerate the discovery of new chemical structures. For FCD/Test, GMT-PE-SMILES performs best among all models with 19.86\%, while GMT-SEMILES and GMT-SELFIES have values 
with72.94\% and 377.5\%. GMT-SMILES, GMT-PE-SMILES and GMT-SELFIES also achieve high value with 16.50\%, 10.87\% and 10.96\%, respectively.

\begin{table}[]
\caption{Comparison of the MOSES Benchmarking Results}
\centering
\label{table:performance}
\begin{tabular}{|lll|lll|llll|}
\hline
\multicolumn{3}{|l|}{}                                                                           & \multicolumn{3}{c|}{GMT}                                                                                                                                                                                             & \multicolumn{4}{c|}{MOSES   reference models}                                                                                                                                                                                                                                                                            \\ \cline{4-10} 
\multicolumn{3}{|l|}{\multirow{-2}{*}{}}                                                         & \multicolumn{1}{l|}{\begin{tabular}[c]{@{}l@{}}GMT-\\ SMILES\end{tabular}} & \multicolumn{1}{l|}{\begin{tabular}[c]{@{}l@{}}GMT-PE-\\ SEMILES\end{tabular}} & \begin{tabular}[c]{@{}l@{}}GMT-\\ SELFIES\end{tabular} & \multicolumn{1}{l|}{\begin{tabular}[c]{@{}l@{}}GCT\\ -SGDR\end{tabular}} & \multicolumn{1}{l|}{VAE}                                                               & \multicolumn{1}{l|}{AAE}                                                       & \begin{tabular}[c]{@{}l@{}}char\\ RNN\end{tabular}                  \\ \hline
\multicolumn{1}{|l|}{validity}               & \multicolumn{1}{l|}{$\uparrow$}                   &        & \multicolumn{1}{l|}{0.8587}                                                & \multicolumn{1}{l|}{0.8288}                                                    & \textbf{1.000}                                         & \multicolumn{1}{l|}{0.9916}                                              & \multicolumn{1}{l|}{\begin{tabular}[c]{@{}l@{}}0.9767$\pm$\\ 0.0012\end{tabular}}          & \multicolumn{1}{l|}{\begin{tabular}[c]{@{}l@{}}0.9368$\pm$\\  0.0341\end{tabular}} & \begin{tabular}[c]{@{}l@{}}0.9748±\\  0.0264\end{tabular}           \\ \hline
\multicolumn{1}{|l|}{unique@1k}              & \multicolumn{1}{l|}{$\uparrow$}                   &        & \multicolumn{1}{l|}{\textbf{1.0000}}                                       & \multicolumn{1}{l|}{\textbf{1.0000}}                                           & \textbf{1.0000}                                        & \multicolumn{1}{l|}{0.998}                                               & \multicolumn{1}{l|}{\textbf{1.0±0.0}}                                                  & \multicolumn{1}{l|}{\textbf{1.0±0.0}}                                          & \textbf{1.0±0.0}                                                    \\ \hline
\multicolumn{1}{|l|}{unique@10k}             & \multicolumn{1}{l|}{$\uparrow$}                   &        & \multicolumn{1}{l|}{0.9998}                                                & \multicolumn{1}{l|}{0.9995}                                                    & \textbf{1.0000}                                        & \multicolumn{1}{l|}{0.9797}                                              & \multicolumn{1}{l|}{\begin{tabular}[c]{@{}l@{}}0.9984±\\ 0.0005\end{tabular}}          & \multicolumn{1}{l|}{\begin{tabular}[c]{@{}l@{}}0.9973±\\ 0.002\end{tabular}}   & \begin{tabular}[c]{@{}l@{}}0.9994 ±\\  0.0003\end{tabular}          \\ \hline
\multicolumn{1}{|l|}{IntDive}                & \multicolumn{1}{l|}{$\uparrow$}                   &        & \multicolumn{1}{l|}{0.8569}                                                & \multicolumn{1}{l|}{0.8558}                                                    & \textbf{0.8701}                                        & \multicolumn{1}{l|}{0.8458}                                              & \multicolumn{1}{l|}{\begin{tabular}[c]{@{}l@{}}0.8558±\\ 0.0004\end{tabular}}          & \multicolumn{1}{l|}{\begin{tabular}[c]{@{}l@{}}0.8557±\\ 0.0031\end{tabular}}  & \begin{tabular}[c]{@{}l@{}}0.8562 ±\\  0.0005\end{tabular}          \\ \hline
\multicolumn{1}{|l|}{filters}                & \multicolumn{1}{l|}{$\uparrow$}                   &        & \multicolumn{1}{l|}{0.9766}                                                & \multicolumn{1}{l|}{0.9797}                                                    & 0.7961                                                 & \multicolumn{1}{l|}{\textbf{0.9982}}                                     & \multicolumn{1}{l|}{\begin{tabular}[c]{@{}l@{}}0.6949±\\ 0.0069\end{tabular}}          & \multicolumn{1}{l|}{\begin{tabular}[c]{@{}l@{}}0.9960±\\ 0.0006\end{tabular}}  & \begin{tabular}[c]{@{}l@{}}0.9943 ±\\  0.0034\end{tabular}          \\ \hline
\multicolumn{1}{|l|}{novelty}                & \multicolumn{1}{l|}{$\uparrow$}                   &        & \multicolumn{1}{l|}{0.9531}                                                & \multicolumn{1}{l|}{0.8829}                                                    & \textbf{0.9683}                                        & \multicolumn{1}{l|}{0.6756}                                              & \multicolumn{1}{l|}{\begin{tabular}[c]{@{}l@{}}0.6949±\\ 0.0069\end{tabular}}          & \multicolumn{1}{l|}{\begin{tabular}[c]{@{}l@{}}0.7931±\\ 0.0285\end{tabular}}  & \begin{tabular}[c]{@{}l@{}}0.8419 ±\\  0.0509\end{tabular}          \\ \hline
\multicolumn{1}{|l|}{}                       & \multicolumn{1}{l|}{}                    & Test   & \multicolumn{1}{l|}{0.5381}                                                & \multicolumn{1}{l|}{0.5778}                                                    & 0.4673                                                 & \multicolumn{1}{l|}{\textbf{0.6513}}                                     & \multicolumn{1}{l|}{\begin{tabular}[c]{@{}l@{}}0.6257±\\ 0.0005\end{tabular}}          & \multicolumn{1}{l|}{\begin{tabular}[c]{@{}l@{}}0.6081±\\ 0.0043\end{tabular}}  & \begin{tabular}[c]{@{}l@{}}0.6015 ±\\  0.0206\end{tabular}          \\ \cline{3-10} 
\multicolumn{1}{|l|}{\multirow{-2}{*}{SNN}}  & \multicolumn{1}{l|}{\multirow{-2}{*}{↑}} & TestSF & \multicolumn{1}{l|}{0.5143}                                                & \multicolumn{1}{l|}{0.5460}                                                    & 0.4485                                                 & \multicolumn{1}{l|}{\textbf{0.5990}}                                     & \multicolumn{1}{l|}{\begin{tabular}[c]{@{}l@{}}0.5783±\\ 0.0008\end{tabular}}          & \multicolumn{1}{l|}{\begin{tabular}[c]{@{}l@{}}0.5677±\\ 0.0045\end{tabular}}  & \begin{tabular}[c]{@{}l@{}}0.5649 ±\\  0.0142\end{tabular}          \\ \hline
\multicolumn{1}{|l|}{}                       & \multicolumn{1}{l|}{}                    & Test   & \multicolumn{1}{l|}{0.7294}                                                & \multicolumn{1}{l|}{\cellcolor[HTML]{FFFFFF}\textbf{0.1986}}                   & 3.7750                                                 & \multicolumn{1}{l|}{0.7980}                                              & \multicolumn{1}{l|}{\begin{tabular}[c]{@{}l@{}}0.0990±\\ 0.0125\end{tabular}}          & \multicolumn{1}{l|}{\begin{tabular}[c]{@{}l@{}}0.5555±\\ 0.2033\end{tabular}}  & \textbf{\begin{tabular}[c]{@{}l@{}}0.0732 ±\\  0.0247\end{tabular}} \\ \cline{3-10} 
\multicolumn{1}{|l|}{\multirow{-2}{*}{FCD}}  & \multicolumn{1}{l|}{\multirow{-2}{*}{↓}} & TestSF & \multicolumn{1}{l|}{1.2607}                                                & \multicolumn{1}{l|}{0.7595}                                                    & 4.5698                                                 & \multicolumn{1}{l|}{0.9949}                                              & \multicolumn{1}{l|}{\begin{tabular}[c]{@{}l@{}}0.5670±\\ 0.0338\end{tabular}}          & \multicolumn{1}{l|}{\begin{tabular}[c]{@{}l@{}}1.0572±\\ 0.2375\end{tabular}}  & \textbf{\begin{tabular}[c]{@{}l@{}}0.5204 ±\\  0.0379\end{tabular}} \\ \hline
\multicolumn{1}{|l|}{}                       & \multicolumn{1}{l|}{}                    & Test   & \multicolumn{1}{l|}{0.9879}                                                & \multicolumn{1}{l|}{0.9982}                                                    & 0.9869                                                 & \multicolumn{1}{l|}{0.9922}                                              & \multicolumn{1}{l|}{\begin{tabular}[c]{@{}l@{}}0.9994±\\ 0.0001\end{tabular}}          & \multicolumn{1}{l|}{\begin{tabular}[c]{@{}l@{}}0.9910±\\ 0.0051\end{tabular}}  & \textbf{\begin{tabular}[c]{@{}l@{}}0.9998 ±\\  0.0002\end{tabular}} \\ \cline{3-10} 
\multicolumn{1}{|l|}{\multirow{-2}{*}{Frag}} & \multicolumn{1}{l|}{\multirow{-2}{*}{↑}} & TestSF & \multicolumn{1}{l|}{0.9850}                                                & \multicolumn{1}{l|}{0.9958}                                                    & 0.9831                                                 & \multicolumn{1}{l|}{0.8562}                                              & \multicolumn{1}{l|}{\textbf{\begin{tabular}[c]{@{}l@{}}0.9984±\\ 0.0003\end{tabular}}} & \multicolumn{1}{l|}{\begin{tabular}[c]{@{}l@{}}0.9905±\\ 0.0039\end{tabular}}  & \begin{tabular}[c]{@{}l@{}}0.9983 ±\\  0.0003\end{tabular}          \\ \hline
\multicolumn{1}{|l|}{}                       & \multicolumn{1}{l|}{}                    & Test   & \multicolumn{1}{l|}{0.8661}                                                & \multicolumn{1}{l|}{0.9125}                                                    & 0.8431                                                 & \multicolumn{1}{l|}{0.8562}                                              & \multicolumn{1}{l|}{\textbf{\begin{tabular}[c]{@{}l@{}}0.9386±\\ 0.0021\end{tabular}}} & \multicolumn{1}{l|}{\begin{tabular}[c]{@{}l@{}}0.9022±\\ 0.0375\end{tabular}}  & \begin{tabular}[c]{@{}l@{}}0.9242 ±\\  0.0058\end{tabular}          \\ \cline{3-10} 
\multicolumn{1}{|l|}{\multirow{-2}{*}{Scaf}} & \multicolumn{1}{l|}{\multirow{-2}{*}{↑}} & TestSF & \multicolumn{1}{l|}{\textbf{0.1650}}                                       & \multicolumn{1}{l|}{0.1087}                                                    & 0.1096                                                 & \multicolumn{1}{l|}{0.0551}                                              & \multicolumn{1}{l|}{\begin{tabular}[c]{@{}l@{}}0.0588±\\ 0.0095\end{tabular}}          & \multicolumn{1}{l|}{\begin{tabular}[c]{@{}l@{}}0.0789±\\ 0.009\end{tabular}}   & \begin{tabular}[c]{@{}l@{}}0.1101 ±\\  0.0081\end{tabular}          \\ \hline
\end{tabular}
\end{table}

We also train five GMT models using different representations and tokens, generate 30,000 hypothetical molecules and evaluate them using MOSES benchmarking metrics. Table \ref{table:GMTmodels} shows the performance of the comparison of the MOSES Benchmarking Results. All models perform very well in terms of uniqueness, in the range of 99.5\%-100\%. In terms of the novelty of the hypothetical molecules, GMT-PE-SMILES achieves 88.92\%, while all other models exceed 90\%. GMT-PE-SMILES outperforms the other models by a wide margin on FCD/Test at 19.86\%.

\begin{table}[]
\caption{Comparison of the GMT models Results}
\centering
\label{table:GMTmodels}
\centering
\begin{tabular}{|lll|lllll|}
\hline
\multicolumn{3}{|l|}{}                                                                           & \multicolumn{5}{c|}{GMT models}                                                                                                                                                                                                                                                                                                                                               \\ \cline{4-8} 
\multicolumn{3}{|l|}{\multirow{-2}{*}{}}                                                         & \multicolumn{1}{l|}{\begin{tabular}[c]{@{}l@{}}GMT-\\ SMILES\end{tabular}} & \multicolumn{1}{l|}{\begin{tabular}[c]{@{}l@{}}GMT-PE-\\ SEMILES\end{tabular}} & \multicolumn{1}{l|}{\begin{tabular}[c]{@{}l@{}}GMT-\\ SELFIES\end{tabular}} & \multicolumn{1}{l|}{\begin{tabular}[c]{@{}l@{}}GMT-\\ DEEP\end{tabular}} & \begin{tabular}[c]{@{}l@{}}GMT-PE-\\ DEEP\end{tabular} \\ \hline
\multicolumn{1}{|l|}{validity}               & \multicolumn{1}{l|}{↑}                   &        & \multicolumn{1}{l|}{\cellcolor[HTML]{FFFFFF}0.8586}                        & \multicolumn{1}{l|}{\cellcolor[HTML]{FFFFFF}0.8288}                            & \multicolumn{1}{l|}{\cellcolor[HTML]{FFFFFF}\textbf{1.0000}}                & \multicolumn{1}{l|}{0.8168}                                              & 0.7954                                                 \\ \hline
\multicolumn{1}{|l|}{unique@1k}              & \multicolumn{1}{l|}{↑}                   &        & \multicolumn{1}{l|}{\cellcolor[HTML]{FFFFFF}\textbf{1.0000}}               & \multicolumn{1}{l|}{\cellcolor[HTML]{FFFFFF}\textbf{1.0000}}                   & \multicolumn{1}{l|}{\cellcolor[HTML]{FFFFFF}\textbf{1.0000}}                & \multicolumn{1}{l|}{\textbf{1.0000}}                                     & \textbf{1.0000}                                        \\ \hline
\multicolumn{1}{|l|}{unique@10k}             & \multicolumn{1}{l|}{↑}                   &        & \multicolumn{1}{l|}{\cellcolor[HTML]{FFFFFF}0.9998}                        & \multicolumn{1}{l|}{\cellcolor[HTML]{FFFFFF}0.9995}                            & \multicolumn{1}{l|}{\cellcolor[HTML]{FFFFFF}\textbf{1.0000}}                & \multicolumn{1}{l|}{\textbf{1.0000}}                                     & 0.9997                                                 \\ \hline
\multicolumn{1}{|l|}{IntDive}                & \multicolumn{1}{l|}{↑}                   &        & \multicolumn{1}{l|}{\cellcolor[HTML]{FFFFFF}0.8569}                        & \multicolumn{1}{l|}{\cellcolor[HTML]{FFFFFF}0.8558}                            & \multicolumn{1}{l|}{\cellcolor[HTML]{FFFFFF}\textbf{0.8701}}                & \multicolumn{1}{l|}{0.8570}                                              & 0.8519                                                 \\ \hline
\multicolumn{1}{|l|}{filters}                & \multicolumn{1}{l|}{↑}                   &        & \multicolumn{1}{l|}{\cellcolor[HTML]{FFFFFF}0.9765}                        & \multicolumn{1}{l|}{\cellcolor[HTML]{FFFFFF}0.9797}                            & \multicolumn{1}{l|}{\cellcolor[HTML]{FFFFFF}0.7961}                         & \multicolumn{1}{l|}{0.9844}                                              & \textbf{0.9847}                                        \\ \hline
\multicolumn{1}{|l|}{novelty}                & \multicolumn{1}{l|}{↑}                   &        & \multicolumn{1}{l|}{\cellcolor[HTML]{FFFFFF}0.9532}                        & \multicolumn{1}{l|}{\cellcolor[HTML]{FFFFFF}0.8829}                            & \multicolumn{1}{l|}{\cellcolor[HTML]{FFFFFF}\textbf{0.9683}}                & \multicolumn{1}{l|}{0.9367}                                              & 0.9149                                                 \\ \hline
\multicolumn{1}{|l|}{}                       & \multicolumn{1}{l|}{}                    & Test   & \multicolumn{1}{l|}{\cellcolor[HTML]{FFFFFF}0.5381}                        & \multicolumn{1}{l|}{\cellcolor[HTML]{FFFFFF}\textbf{0.5778}}                   & \multicolumn{1}{l|}{\cellcolor[HTML]{FFFFFF}0.4673}                         & \multicolumn{1}{l|}{0.5509}                                              & 0.5722                                                 \\ \cline{3-8} 
\multicolumn{1}{|l|}{\multirow{-2}{*}{SNN}}  & \multicolumn{1}{l|}{\multirow{-2}{*}{↑}} & TestSF & \multicolumn{1}{l|}{\cellcolor[HTML]{FFFFFF}0.5143}                        & \multicolumn{1}{l|}{\cellcolor[HTML]{FFFFFF}\textbf{0.5460}}                   & \multicolumn{1}{l|}{\cellcolor[HTML]{FFFFFF}0.4485}                         & \multicolumn{1}{l|}{0.5246}                                              & 0.5405                                                 \\ \hline
\multicolumn{1}{|l|}{}                       & \multicolumn{1}{l|}{}                    & Test   & \multicolumn{1}{l|}{\cellcolor[HTML]{FFFFFF}0.7294}                        & \multicolumn{1}{l|}{\cellcolor[HTML]{FFFFFF}\textbf{0.1986}}                   & \multicolumn{1}{l|}{\cellcolor[HTML]{FFFFFF}3.7750}                         & \multicolumn{1}{l|}{0.3604}                                              & 0.4366                                                 \\ \cline{3-8} 
\multicolumn{1}{|l|}{\multirow{-2}{*}{FCD}}  & \multicolumn{1}{l|}{\multirow{-2}{*}{↓}} & TestSF & \multicolumn{1}{l|}{\cellcolor[HTML]{FFFFFF}1.2607}                        & \multicolumn{1}{l|}{\cellcolor[HTML]{FFFFFF}\textbf{0.7595}}                   & \multicolumn{1}{l|}{\cellcolor[HTML]{FFFFFF}4.5698}                         & \multicolumn{1}{l|}{0.9563}                                              & 1.0736                                                 \\ \hline
\multicolumn{1}{|l|}{}                       & \multicolumn{1}{l|}{}                    & Test   & \multicolumn{1}{l|}{\cellcolor[HTML]{FFFFFF}0.9879}                        & \multicolumn{1}{l|}{\cellcolor[HTML]{FFFFFF}\textbf{0.9982}}                   & \multicolumn{1}{l|}{\cellcolor[HTML]{FFFFFF}0.9869}                         & \multicolumn{1}{l|}{0.9981}                                              & 0.9967                                                 \\ \cline{3-8} 
\multicolumn{1}{|l|}{\multirow{-2}{*}{Frag}} & \multicolumn{1}{l|}{\multirow{-2}{*}{↑}} & TestSF & \multicolumn{1}{l|}{\cellcolor[HTML]{FFFFFF}0.9850}                        & \multicolumn{1}{l|}{\cellcolor[HTML]{FFFFFF}0.9958}                            & \multicolumn{1}{l|}{\cellcolor[HTML]{FFFFFF}0.9831}                         & \multicolumn{1}{l|}{\textbf{0.9964}}                                     & 0.9934                                                 \\ \hline
\multicolumn{1}{|l|}{}                       & \multicolumn{1}{l|}{}                    & Test   & \multicolumn{1}{l|}{\cellcolor[HTML]{FFFFFF}0.8661}                        & \multicolumn{1}{l|}{\cellcolor[HTML]{FFFFFF}\textbf{0.9125}}                   & \multicolumn{1}{l|}{\cellcolor[HTML]{FFFFFF}0.8431}                         & \multicolumn{1}{l|}{0.8880}                                              & 0.8903                                                 \\ \cline{3-8} 
\multicolumn{1}{|l|}{\multirow{-2}{*}{Scaf}} & \multicolumn{1}{l|}{\multirow{-2}{*}{↑}} & TestSF & \multicolumn{1}{l|}{\cellcolor[HTML]{FFFFFF}\textbf{0.1649}}               & \multicolumn{1}{l|}{\cellcolor[HTML]{FFFFFF}0.1087}                            & \multicolumn{1}{l|}{\cellcolor[HTML]{FFFFFF}0.1096}                         & \multicolumn{1}{l|}{0.1511}                                              & 0.1170                                                 \\ \hline
\end{tabular}
\end{table}

\FloatBarrier

\begin{figure}[ht]
  \centering
  \includegraphics[width=0.9\linewidth]{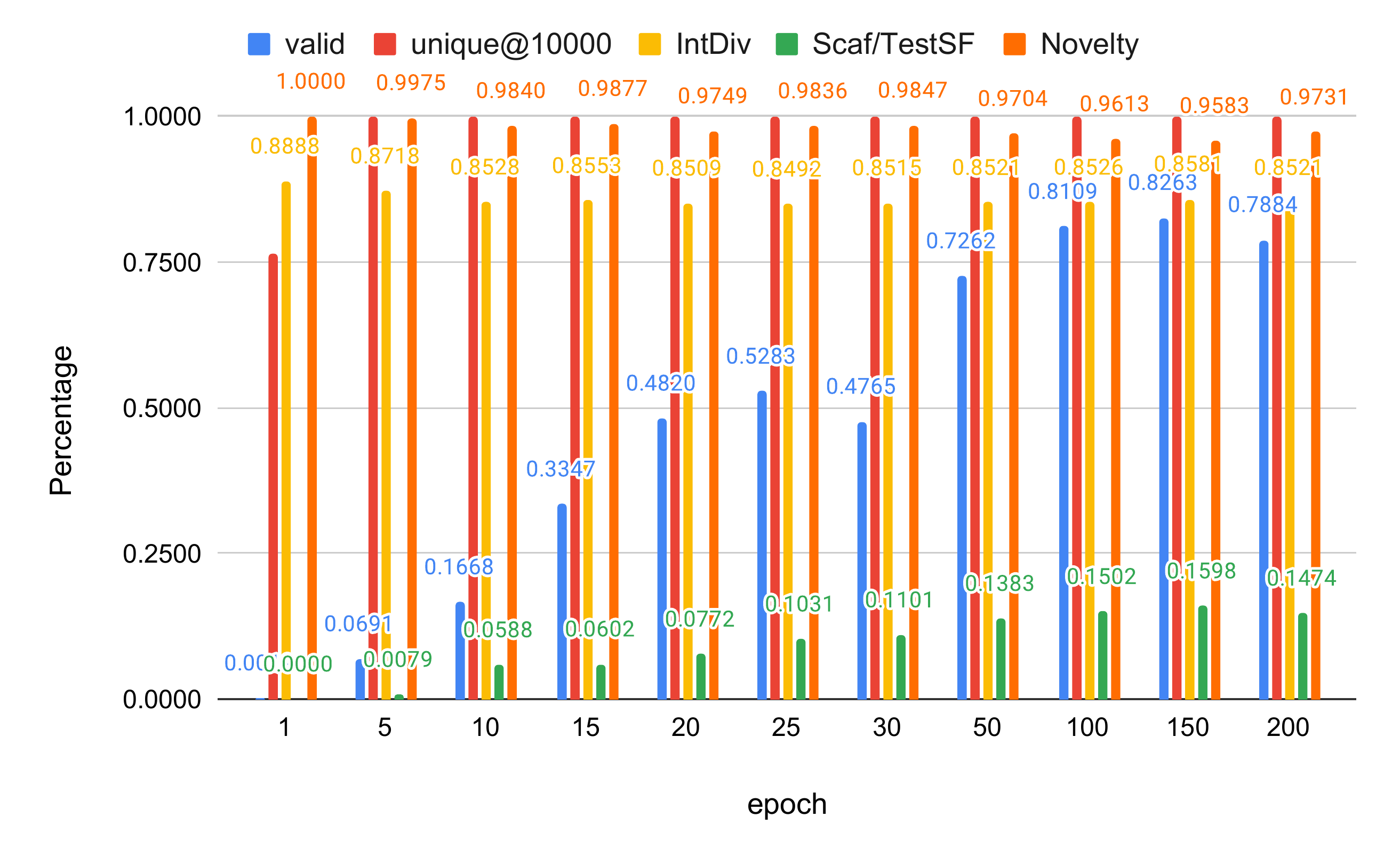}
  \caption{Percentages of valid, unique@10000, intDiv and Scaf/TestSF samples generated by the SMILES atom tokenizer models saved over the training process. The models generate few valid SMILES strings in the beginning. As the training goes on, the models gradually gain the capability to generate chemically valid SMILES molecules compositions.}
  \label{fig:SMILESprocess}
\end{figure}

\begin{figure}[ht]
  \centering
  \includegraphics[width=0.9\linewidth]{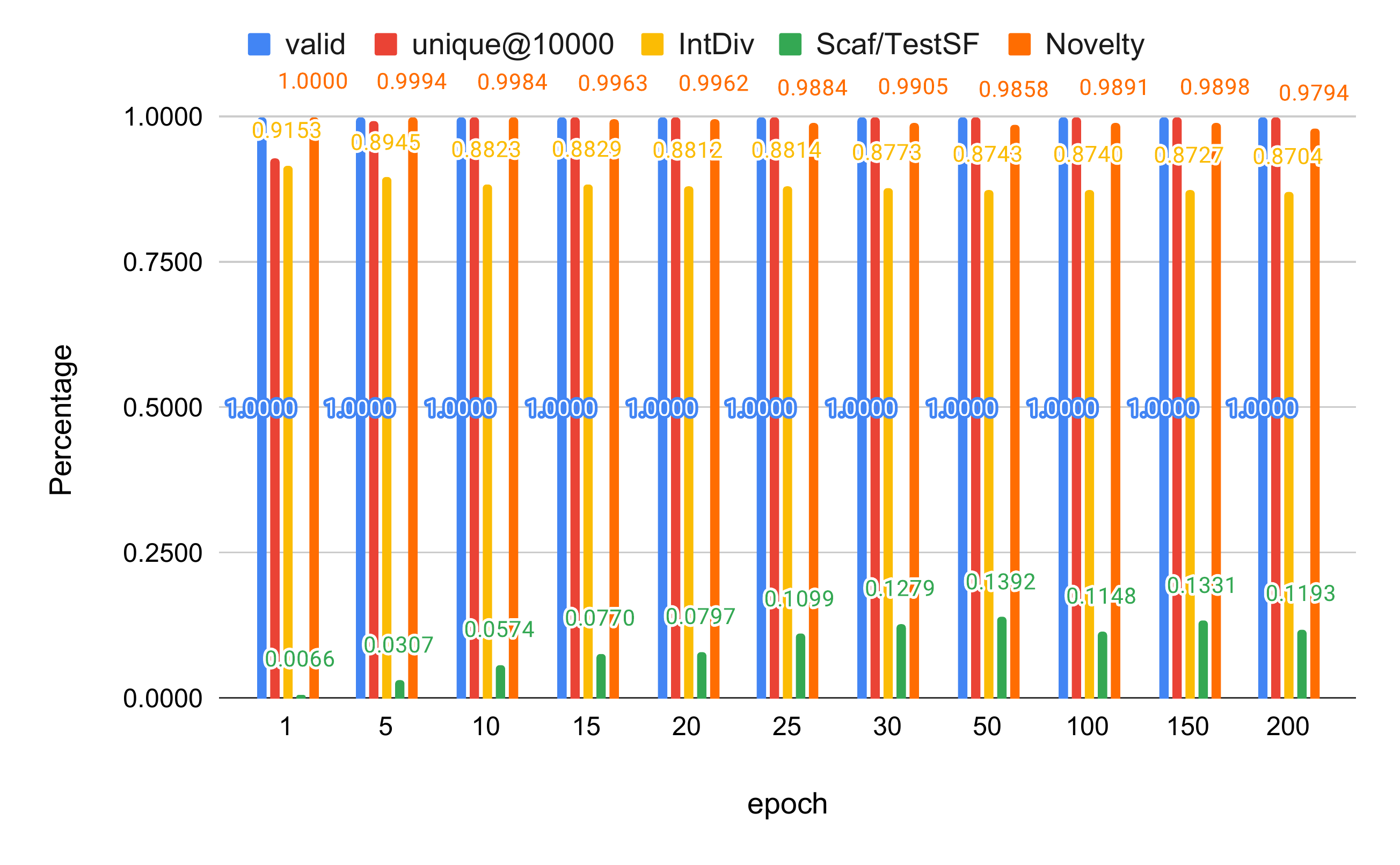}
  \caption{Percentages of valid, unique@10000, intDiv and Scaf/TestSF samples generated by the SELFIES atom tokenizer models saved over the training process. The models generate almost one hundred percent valid SMILES strings from the beginning to the end and the Scaf/TestSF value has also been growing with epoch from 1 to 200.}
  \label{fig:SELFIESprocess}
\end{figure}

\paragraph{Process of GMT's learning of chemical rules:}  To illustrate the chemical order/rules emerge during the training process of our GMT models, we save the intermediate models at the end of 1/5/10/15/20/25/30/50/100/150/200 epochs of training using the SMILES and SELFIES dataset, respectively. Then we use 30,000 generated samples to evaluate validity, unique@10k, IntDiv, Scaf/TestSF and Novelty with MOSES benchmarking metrics. As shown in Figure \ref{fig:SMILESprocess}, the validity of the model using the SMILES representation is only about 50\% of the maximum value when the epoch of the model training is less than 30, and its validity exceeds 80\% at the 100 epochs. This growing process shows that the model is learning the valence rules and the syntax of the SMILES language. 
For the model using SELFIES representation, the results are shown in Figure \ref{fig:SELFIESprocess}. Because every SELFIES syntax is guaranteed to correspond to a valid molecule \cite{krenn2019selfies}, the validity is always 100\% throughout training epochs from 1 to 200. The increase in Scaf/TestSF value also indicates that the model has learned the Bemis–Murcko scaffold \cite{bemis1996properties}, which contains all molecule’s ring structures and linker fragments connecting rings. 

\FloatBarrier

\FloatBarrier

\subsection{Comparison of different molecule representations: SMILES, SELFIES, and DeepSMILES }

Different representations make the model more capable of generating new potential molecules. We use three types of string-based molecular representations: The simplified molecular input line entry system (SMILES) \cite{weininger1988smiles}, SELF-referencIng Embedded Strings (SELFIES) \cite{krenn2020self}, DeepSMILES \cite{o2018deepsmiles} and two kinds of tokenizers: Atom-level and SmilesPE \cite{li2021smiles}. Table \ref{table:representations} shows examples of the different molecule representations with two types of tokenizers. SELFIES only has the atom-level tokenizers. We first use SMILES, which is the most widely used representation in computational chemistry. SMILES has some weaknesses such as multiple different SMILES strings can represent the same molecule and it is not robust because it is possible for generative models to create strings that do not represent valid molecular graphs. DeepSMILES is a modification of SMILES which obviates most syntactic errors, while semantic mistakes were still possible \cite{o2018deepsmiles}. Therefore, we also use the representation of SELFIES, which can generate 100\% effective molecular graph to definitely avoid the problem of model robustness. SELFIES is like an automaton or derivation grammar, which is designed to eliminate syntactic and semantic invalid strings. Atomic-level tokenization is a method commonly used in deep learning, which simply breaks the SMILES string character-by-character, with each character serving as a token. We use not only an atom-level tokenizer, but also the SmilesPE representation, which has shorter input sequences and can save the computational cost of model training and inference. SmilesPE identifies and retains frequent SMILES substrings as unique tokens, where each token is represented as a chemically meaningful substructure.

\begin{table}[ht]
\centering
\caption{Comparison of the different molecule representations: SMILES, SELFIES, and
DeepSMILE}
\label{table:representations}
\begin{tabular}{|l|l|}
\hline
\rowcolor[HTML]{FFFFC7} 
Tokenizer  & Atom-level                                                                                       \\ \hline
SMILES     & C O c 1 c c c c c 1 O C ( = O ) O c 1 c c c c c 1 O C                                            \\ \hline
DeepSMILES & C O c c c c c c 6 O C = O ) O c c c c c c 6 O C                                                  \\ \hline
SELFIES    & {[}C{]} {[}N{]} {[}C{]} {[}Branch1{]} {[}C{]} {[}P{]} {[}C{]} {[}C{]} {[}Ring1{]} {[}=Branch1{]} \\ \hline
\rowcolor[HTML]{FFFFC7} 
Tokenizer  & SmilesPE                                                                                         \\ \hline
SMILES     & COc1ccccc1 O C(=O)O c1ccccc1 OC                                                                  \\ \hline
DeepSMILES & CO cccc cc 6 OC =O) O cccc cc 6 OC                                                               \\ \hline
\end{tabular}
\end{table}

\subsection{Conditional generative design of molecules}

\begin{table}[ht]
\centering
\caption{Datasets for conditional generation}
\begin{tabular}{|
>{\columncolor[HTML]{FFFFC7}}c |c|c|c|}
\hline
\multicolumn{1}{|l|}{\cellcolor[HTML]{FFFFC7}{\color[HTML]{000000} }} & \cellcolor[HTML]{FFFFC7}{\color[HTML]{000000} Whole Set} & \cellcolor[HTML]{FFFFC7}{\color[HTML]{000000} Training Set (Top 50\%)} & \cellcolor[HTML]{FFFFC7}{\color[HTML]{000000} Generated Samples} \\ \hline
LogP                                                                  & 1,584,662                                                & 792,331                                                                & 16,748                                                           \\ \hline
tPSA                                                                  & 1,584,662                                                & 792,331                                                                & 16,643                                                           \\ \hline
QED                                                                   & 1,584,662                                                & 792,331                                                                & 17,082                                                           \\ \hline
\end{tabular}
\end{table}

\begin{figure}[ht]
\begin{subfigure}[t]{0.5\textwidth}
        \includegraphics[width=\textwidth]{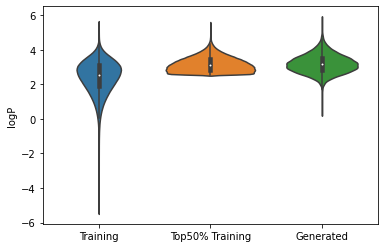}
        \caption{logP}
        \vspace{-3pt}
        \label{fig:logP}
    \end{subfigure}
    \begin{subfigure}[t]{0.5\textwidth}
        \includegraphics[width=\textwidth]{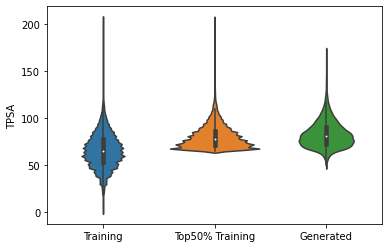}
        \caption{tPSA}
        \vspace{-3pt}
        \label{fig:TPSA}
    \end{subfigure} 
 \begin{subfigure}[t]{0.50\textwidth}
        \includegraphics[width=\textwidth]{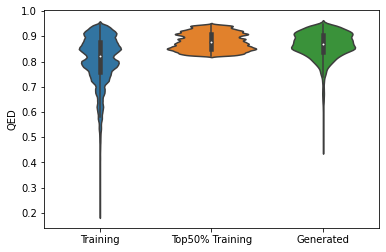}
        \caption{QED}
        \vspace{-3pt}
        \label{fig:QED}
    \end{subfigure}              
\caption{Comparison of property distribution of three different datasets: the whole MOSES training set, the top 50\% properties set used for training the conditional generator models, and the generated samples set for logP, tPSA, QED.}

  \label{fig:distributionmol}
\end{figure}

One desirable generation capability of molecular generators is to design molecules that optimize one or more specific properties. Here we test our model with such capability by building three conditional generators. Basically, we prepare three different training sets from the MOSES training dataset by picking samples whose corresponding property values are within the top 50\% of the whole MOSES training dataset, where the three properties include the octanol-water partition coefficient (logP), the topological Polar Surface Area (tPSA), and the Quantitative Estimate of Drug-likeness (QED), which are computed using the RDKit . We then train the generators with these high-property training molecules and use them to generate 20,000 candidate samples, which are then fed to the RDkit for the property calculation. It is found that RDKit cannot calculate the properties for some of these generated samples. After filtering these generated samples, we finally obtain 16,748, 16,643, and 17,082 samples for LogP, tPSA, and QED respectively. The distributions of these properties values of the whole dataset, the biased (top 50\%) training set, and the generated candidate sets are shown in Figure \ref{fig:distributionmol}. 
It is found that for all three properties, the distributions of our generated molecules are much closer to those of the top 50\% training sets compared to the property distributions of the whole MOSES training dataset, which indicates that the GMTransformer models have learned the implicit rules to generate high-property molecules.

\FloatBarrier

\section{Discussion}

We propose Generative Molecular Transformer (GMTransformer), a probabilistic generative language model based on neural networks and transformers for the generation and design of molecules. The advantages of the GMT model lie in its interpretability and data efficiency as shown in our previous work on generating hypothetical inorganic materials\cite{wei2022crystal}. Here we show that it can rapidly learn the grammar rules of molecular graphs and generate high-quality hypothetical molecules. Our model is based on the blank filling language model which has a unique advantage and potential for tinkering molecule design as we showed in \cite{wei2022crystal} for tinkering design of materials compositions.

In order to evaluate the performance of our models in generating valid and potential molecules, we compare it with a baseline language-based model and three other deep neural network models using a selected set of criteria from the Moses metrics. The results of the hypothetical molecules generated by our models and the high-property distribution results show that our GMT models are able to learn the grammar rules of molecules to generate valid molecules. As shown in Table \ref{table:performance}, our GMT-SMILES, GMT-PE-SMILES and GMT-SELFIES generate 85.87\%, 82.88\% and 100\% valid samples, respectively. Especially, the novelty of GMT-SMILES, GMT-PE-SMILES and GMT-SELFIES is as high as 95.31\%, 88.29\% and 96.83\% respectively compared to 67.56\% of the baseline GCT-SGDR. At the same time, GMT-SMILES, GMT-PE-SMILES and GMT-SELFIES have the highest IntDive scores with 85.69\%, 85.58\% and 87.01\% respectively among all benchmarked models, which means that they can generate samples with higher diversity with the potential to accelerate the discovery of new chemical structures.

While uniqueness, validity, and novelty are evaluated mainly based on the molecule structure, the relevance of generated samples to druggability and biological processes are not clear. To address this issue, we evaluate our models using the FCD criterion \cite{preuer2018frechet}, which is computed using the activation of the penultimate layer of ChemNet. This criterion can capture both chemical and biological property of the generated molecules. We find that out of the six language models (in Table \ref{table:performance} and \ref{table:GMTmodels}), our GMT-PE-SMILES achieves the best performance in terms of the FCD/Test measure with 19.86\%, while GMT-SMILES shows the performance with 72.94\% and the baseline GCT-SGDR shows 79.80\% of the FCD/Test. However, the FCD/Test performance of the GMT-SELFIES model is relatively low without clear reason. 
We also find the FCD performance is also relatively low in other relevant models \cite{yang2022exploring,gnaneshwar2022score} that also uses SELFIES representation. 

Another advantage of our GMTransformer for molecule generation is that it allows to use functional groups of molecules as tokens to train models that generate molecules with specific functions. While fragment-based models have been proposed before, the blank filling model we use here can be used to discover those function groups as highly dependent subsequences. The discovery and usage of these special functional groups of molecules may have great potential for molecule design for specific functions suitable for real-life scenarios \cite{mowbray2008influence}. We also find that the molecule sequence rewriting probabilities and interpretability of the GMT model provide more control over the molecular generation process, which brings more potential for generating molecules with specific properties. This has been demonstrated in our materials composition design using the BLM model \cite{wei2022crystal}. 

\section{Materials and Methods}
\label{sec:others}

\subsection{Datasets}

We use the dataset from the benchmarking platform Molecular Sets (MOSES) at \url{https://github. com/molecularsets/moses} \cite{polykovskiy2020molecular}. It contains 1,936,962 molecular structures totally and splits them into three datasets for experiments. Each of them consists of training samples (around 1.6 M), test samples (176 k), and scaffold test samples (176 k) and we use the training and test sets in our experiments. We use the SMILES, SELFIES sets with the basic Atom-level and SmilesPE tokenizers.

\subsection{Evaluation criteria}
We use the MOSES benchmarking score metrics to evaluate the overall quality of the generated samples.
Several models with different tokens are used for GMTransformer training and each model generates 30,000 samples that are evaluated by the MOSES benchmarking metrics in Table \ref{table:performance}. 
The ratio of valid and unique (unique@$1k$ and unique@$10k$) are reporting the validity and uniqueness of the generated SMILES string respectively. Novelty is the proportion of molecules in the generated samples that is not in the training set. Filter refers to the proportion of generated molecules that passed the filter during dataset construction. The MOSES metrics also measure the internal diversity (IntDiv) \cite{benhenda2018can}, the similarity to the nearest neighbor (SNN) \cite{polykovskiy2020molecular}, Frechet ChemNet distance (FCD) \cite{ preuer2018frechet}, fragment similarity (Frag) \cite{polykovskiy2020molecular}, and scaffold similarity (Scaf) \cite{polykovskiy2020molecular}.

The Internal diversity (IntDiv) is calculated via eq (\ref{eq:intdiv}), it evaluates the chemical diversity in the generated set $G$ of molecules and detects if the generative model has model collapse. 

\begin{equation}
\operatorname{IntDiv{_p}}(G) = 1-\sqrt[p]{\frac{1}{|G|^{2}} \sum_{{m_{1}, m_{2}} \in G} \operatorname{T} {\left(m_{1}, m_{2}\right)}^{p}}
\label{eq:intdiv}
\end{equation}

Where $G$ is the generated set, $m_a$ and $m_b$ are their Morgan fingerprints \cite{rogers2010extended} for two molecules $a$ and $b$. $T$ is the Tanimoto-distance \cite{tanimoto1958elementary} molecules of generated set $G$.

The Similarity to a nearest neighbor (SNN) is calculated via eq (\ref{eq:SNN}). 

\begin{equation}
\operatorname{SNN}(G, R)=\frac{1}{|G|} \sum_{m_{G} \in G} \max _{m_{R} \in R} T\left(m_{G}, m_{R}\right)
\label{eq:SNN}
\end{equation}

Where $m$ is the Morgan fingerprints of a molecule. T($m_G,m_R$) is an average Tanimoto similarity between $m_G$ in generated set $G$ and its nearest neighbor molecule $m_R$ in the reference dataset $R$.

The Fr\'echet ChemNet distance (FCD) is computed from the activation of the penultimate layer of the deep neural network ChemNet, which was trained to predict the biological activity of drugs. These activations can capture chemical and biological properties of compounds for two sets $G$ and $R$. It is defined as eq \ref{eq:FCD}):

\begin{equation}
\operatorname{FCD}(G, R)=\lvert\lvert\mu_{G}-\mu_{R}\rvert\rvert^{2}+\operatorname{Tr}\left(\sum_{G}+\sum_{R}-2\left(\sum_{G} \sum_{R}\right)^{1 / 2}\right)
\label{eq:FCD}
\end{equation}

Where $\mu_{G}$, $\mu_{R}$ are mean vectors for sets $G$ and $R$ respectively, $\sum{G}$, $\sum{R}$ are full covariance matrices of activations. $Tr$ stands for the trace operator.

The Fragment similarity (Frag) is calculated via eq (\ref{eq:Frag}), which compares distributions of BRICS fragments \cite{degen2008art} in the generated set $G$ and reference set $R$.

\begin{equation}
\operatorname{Frag}(G, R)=\frac{\sum_{f \in F}\left(c_{f}(G) \cdot c_{f}(R)\right)}{\sqrt{\sum_{f \in F} c_{f}^{2}(G)} \sqrt{\sum_{f \in F} c_{f}^{2}(R)}}
\label{eq:Frag}
\end{equation}

Where $F$ is the set of BRICS fragments. $c_{f}(X)$ stands for the frequency of occurrences of a substructure fragment $f$ in the molecules of set $X$.

The Scaffold similarity (Scaff) is similar with Frag but it computes the frequencies of Bemis–Murcko scaffolds \cite{bemis1996properties}. It is calculated as eq (\ref{eq:Scaf}):

\begin{equation}
\operatorname{Scaf}(G, R)=\frac{\sum_{s \in S}\left(c_{s}(G) \cdot c_{s}(R)\right)}{\sqrt{\sum_{s \in S} c_{s}^{2}(G)} \sqrt{\sum_{s \in S} c_{s}^{2}(R)}}
\label{eq:Scaf}
\end{equation}

Where S is the set of Bemis-Murcko scaffolds, $s_{S}(X)$ stands for the frequency of occurrences of a substructure scaffold $s$ in the molecules of set $X$.

is when the number of layers is 15. The values of Validity, Unique@10000, Filters, and Novelty at this point are 86.46\%, 99.99\%, 98.06\%, 94.13\% respectively. For the values of FCD/Test and Scaf/TestSF are 32.08\% and 15.41\% respectively. We use the default number of layers for the model of 6 instead of 15 because hyper-parameter studies show that the number of layers has little effect on the overall performance of the model, and the model with the default number of layers has higher efficiency.

\begin{table}[]
\caption{Hyper-parameter tuning of GMTransform molecules generator}
\label{table:hyperparameter}
\centering
\begin{tabular}{|l|l|l|l|l|l|}
\hline
Number of layers      &        & 5      & 10     & 15     & 20     \\ \hline
valid                 &        & 0.8582 & 0.8488 & 0.8646 & 0.8549 \\ \hline
unique@1000           &        & 1.0000 & 1.0000 & 1.0000 & 1.0000 \\ \hline
unique@10000          &        & 1.0000 & 0.9997 & 0.9999 & 0.9998 \\ \hline
IntDiv                &        & 0.8529 & 0.8536 & 0.8541 & 0.8540 \\ \hline
Filters               &        & 0.9802 & 0.9838 & 0.9806 & 0.9812 \\ \hline
Novelty               &        & 0.9351 & 0.9389 & 0.9413 & 0.9362 \\ \hline
\multirow{2}{*}{SNN}  & Test   & 0.5559 & 0.5556 & 0.5509 & 0.5554 \\ \cline{2-6} 
                      & TestSF & 0.5277 & 0.5279 & 0.5252 & 0.5304 \\ \hline
\multirow{2}{*}{FCD}  & Test   & 0.5404 & 0.3243 & 0.3108 & 0.3903 \\ \cline{2-6} 
                      & TestSF & 1.1415 & 0.8461 & 0.7978 & 0.8609 \\ \hline
\multirow{2}{*}{Frag} & Test   & 0.9939 & 0.9950 & 0.9965 & 0.9966 \\ \cline{2-6} 
                      & TestSF & 0.9904 & 0.9913 & 0.9933 & 0.9950 \\ \hline
\multirow{2}{*}{Scaf} & Test   & 0.8954 & 0.8902 & 0.8955 & 0.8868 \\ \cline{2-6} 
                      & TestSF & 0.1425 & 0.1482 & 0.1541 & 0.1285 \\ \hline
\end{tabular}
\end{table}

\FloatBarrier

\section*{Acknowledgement}

\paragraph{Funding:}The research reported in this work was supported in part by National Science Foundation under the grant and 1940099 and 1905775. The views, perspectives, and content do not necessarily represent the official views of the NSF. QW would like to acknowledge the seed funding support from the Big Data Health Science Center (BDHSC) of the University of South Carolina.
\paragraph{Author contributions:}
Conceptualization, J.H.; methodology,J.H. L.W., N.F., Y.S., Q.W.; software, J.H., L.W.,N.F.; resources, J.H.; writing--original draft preparation, J.H., L.W., N.F.; writing--review and editing, J.H, L.W.; visualization, L.W., Y.S., J.H.; supervision, J.H.;  funding acquisition, J.H.,Q.W.
\paragraph{Competing interests: The authors declare that they have no competing interests}
\paragraph{Data and code availability:}
The raw molecules QM9 dataset is downloaded from http://quantum-machine.org/datasets/. 
The preprocessed datasets and code download link can be found at \url{http://github.com/usccolumbia/GMTransformer}

\bibliographystyle{unsrt}  
\bibliography{references}

\end{document}